# The impact of COVID-19 on the stock market crash risk in China

Zhifeng Liu[1,6], Toan Luu Duc Huynh[3,4,5], Peng-Fei Dai[2*]

[1] Management School, Hainan University, Haikou, China
[2] College of Management and Economics, Tianjin University, Tianjin, China
[3] Institute of Research and Development, Duy Tan University, Danang 550000, Vietnam
[4] Faculty of Business Administration, Duy Tan University, Danang 550000, Vietnam
[5] Chair of Behavioral Finance, WHU – Otto Beisheim School of Management, Vallendar, Germany
[6] Supply Chain and Logistics Optimization Research Center, University of Windsor, Windsor, Canada

**ABSTRACT**

This study investigates the impact of the COVID-19 pandemic on the stock market crash risk in China. For this purpose, we first estimated the conditional skewness of the return distribution from a GARCH with skewness (GARCH-S) model as the proxy for the equity market crash risk of the Shanghai Stock Exchange. We then constructed a fear index for COVID-19 using data from the Baidu Index. Based on the findings, conditional skewness reacts negatively to daily growth in total confirmed cases, indicating that the pandemic increases stock market crash risk. Moreover, the fear sentiment exacerbates such risk, especially with regard to the impact of COVID-19. In other words, when the fear sentiment is high, the stock market crash risk is more strongly affected by the pandemic. Our evidence is robust for the number of daily deaths and global cases.

## 1. Introduction

Due to the onset of the COVID-19 pandemic, there has been a significant decline in stock market prices, which has placed unprecedented pressure on global financial markets. In this regard, existing studies have examined the impact of COVID-19 on either stock market returns or volatility, with some conclusions supported by empirical evidence (Baker et al., 2020; Al-Awadhi et al., 2020; Phan and Narayan, 2020; Ashraf, 2020; Kartal et al., 2020; Ramelli and Wagner, 2020; Zhang, Hu, and Ji, 2020; Sharif, Aloui, and Yarovaya, 2020). In addition, because of the co-movement in global stock markets (Dai et al., 2019; Wen, Yang, and Zhou, 2019, Dai et al., 2020), global equities have plummeted, followed by a spike in market volatility. In a related study, Baker et al. (2020) concluded that the level of market volatility (as of March 2020) could be equivalent to or even surpass previous crises, such as Black Monday (October 1987), the Global Financial Crisis (December 2008), the Great Crash (1929), and the Great Depression (the early 1930s). Schell et al. (2020) also emphasized that this time period is indeed different, implying that only COVID-19 exhibits negative returns from the Public Health Risk Emergency of International Concern (PHEIC) announcements. Thus, motivated by the literature, the present study focuses on another pivotal downside risk during the pandemic: the stock market crash risk and investor sentiment in China. Notwithstanding the literature on infectious disease outbreaks and stock market performance, this study not only quantifies the stock market crash risk in the country but also investigates the role of investor sentiment via the Baidu Index and the number of COVID-19 infected cases and deaths. In doing so, this study sheds light on how COVID-19 proxies and investor behaviors may predict equity market crash risk at the onset of a future pandemic.

After the COVID-19 outbreak, the stock market suffered a severe shock, and stock market crash risk became significantly greater than normal. In fact, during the first three months of 2020 (59 trading days), there were 6 days with a single-day crash of 2% or more. In comparison, over the past three years (730 trading days), there were only 21 days with such a decline. Based on this situation, some scholars have begun to focus on stock market crashes during pandemics. For example, Mazur et al. (2020) discussed COVID-19 and stock market crashes and defined them in terms of extreme returns and volatility.

Crash risk, measured here by conditional skewness, captures negative asymmetry risk and extreme downside risk in the stock market (Chen, Hong, and Stein, 2001).

Previous studies have also analyzed stock market crash risk from different perspectives. For instance, Chen, Hong, and Stein (2001) performed an empirical investigation to forecast crash risk (skewness), both at the firm and whole-market levels, while Kim et al. (2011a; 2011b) focused on crash risk at the firm level.

As for the present study, it investigates the impact of COVID-19 on the crash risk of the Chinese stock market. For this purpose, we first employed the GARCH-S model to estimate the daily time-varying skewness of stock returns and then used it as a measure of stock market crash risk. In the subsequent empirical analysis, we not only analyzed the impact of the severity of the pandemic (measured by the number of daily confirmed cases) on crash risk but also examined the interaction between the severity of the pandemic and investor sentiment.

This study contributes to the literature in several ways. First, we examined the risk of stock market crashes during a pandemic, with specific focus on asymmetric negative and extreme risks. Second, using the Baidu Index, we created a fear sentiment index toward the COVID-19 pandemic to determine whether the panic related to it correlated with stock market crashes. Finally, we investigated the role of fear sentiment with regard to the impact of COVID-19 on stock market crash risk. More importantly, our findings carry several policy implications, including alleviating investor panic to mitigate equity market crash risk and offering a preventive measure related to the number of COVID-19 infected cases and deaths. Our findings may also provide policymakers with a deeper understanding of how to respond to and cope with investor pessimism about equity markets in a timely and comprehensive manner, especially during financial downturns.

The remainder of this study is as follows. Section 2 reviews the current literature, while Section 3 describes the data and methodology. Then, Section 4 summarizes the empirical results regarding the impact of COVID-19 on the stock market crash risk in China. Finally, Section 5 presents the conclusions.

## 2. Literature review

In this study, it is essential to construct a sound theoretical framework on how the COVID-19 pandemic has adversely affected financial markets. Goodell (2020) showed that markets are likely to react similarly to the pandemic as to other disasters, such as natural disasters (Gao et al., 2020) or terrorism (Wang and Young, 2020). There is also a common trait that investors' risk preferences or moods toward certain events might vary considerably, leading to an increase in fear-induced sentiment (He et al., 2019; Liu and Zhang, 2020; He, 2020; Liu et al., 2020; Zhang, Song, and Liu, 2020; Dai et al., 2021). While previous disasters have occurred in specific regions of the world with partial disruptions, the COVID-19 pandemic has disrupted travel as well as economic transactions on a global scale. Hence, the effects of the pandemic on the overall economy will not only significantly influence domestic demand but will also limit supply, negatively impact firms' future cash flows, and foster public pessimism about the future.

The COVID-19 pandemic is considered the most significant global health crisis since the influenza pandemic of 1918. Thus, there are many unknown perspectives to examine, with financial crashes being one of the greatest concerns. Mazur et al. (2020) claimed that the financial market crash of March 2020 was triggered by government reactions. Interestingly, negative effects were more pronounced in specific industries such as the crude oil, real estate, entertainment, and hospitality sectors. Their study also confirmed the findings of Mishkin and White (2002), who found that the equity market crash could result in a drop of 20% to 25% in the United States (U.S.) equity index, compared to previous crises (e.g., World War I, World War II, etc.) due to the sequence of panic selling. Thus, our motivation is to examine the determinants, including investor sentiment.

Previous research has also shown that the pandemic's status may predict equity market crash risk. For instance, Giglio et al. (2020), Wen, Xu, and Ouyang, et al. (2019), Wen, Xu, and Chen, et al. (2019a), and Zhang, Jia, and Chen (2020) showed that short-run investor expectations may correlate with stock market crash risk. It should be noted that previous studies (Giglio et al., 2019; Giglio et al., 2020) also confirmed that the probability of an equity market crash before a crisis is lower because investors tend to be more optimistic about stock market returns. Notwithstanding these findings, a further investigation of COVID-19 is promising, as we take no stance on whether the likelihood of a market crash would significantly change in the two sub-periods, i.e., before and after the pandemic. Given the foregoing discussion and argument, we posited the following hypothesis:

*$H_1$: The equity market crash risk in China during the COVID-19 pandemic is higher than in the preceding period.*

In order to examine $H_1$, we divided our samples into the two aforementioned sub-periods and employed statistical testing. Previous literature (e.g., Giglio et al., 2019; Giglio et al., 2020; Gabaix, 2012; Wachter, 2013) also forms a sound framework for the construction of our second hypothesis, which is posited as follows:

*$H_2$: There is no relationship between investor sentiment and stock market crashes regarding the onset of the COVID-19 pandemic.*

Although there is mounting literature examining how investors have overreacted (or underreacted) to the COVID-19 pandemic (e.g., Aslam et al., 2020; Schell et al., 2020; Yarovaya et al., 2020), what drives the stock market crash risk in China has yet to be addressed. It is marginally relevant to consider that the combination of economic uncertainty and behavioral factors positively contribute to financial asset crash risk (e.g., Bitcoin (Kalyvas et al., 2020) and the Chinese stock market (Jin et al., 2019; Luo and Zhang, 2019; Ju, 2019; Wongchoti et al., 2020)). Notably, how the aforementioned factors drive the stock market during disaster periods remains unclear. Thus, we used two proxies (i.e., the fear index for COVID-19 from the Baidu Index and actual pandemic figures) to predict changes in the stock market crash risk index, constructed via the GARCH-S model. Proxy substitution also served as our alternative approach to determining whether the findings were robust.

Finally, we addressed the gap in the literature via two research questions: 1) What is the equity market crash risk before and after the COVID-19 pandemic? and 2) What drives the stock market crash risk in China, i.e., do investors' fears and/or current COVID-19 statistics matter to the stock market crash risk in China? Addressing these questions will not only benefit practitioners by increasing caution over extreme market shocks but also increase the academic understanding of the empirical evidence. Based on the aforementioned arguments, our research questions are closely related to the literature on financial market reactions to the COVID-19 pandemic. However, few studies have examined the effect of investor sentiment, especially fear sentiment, on systematic risk in emerging economies during the onset of COVID-19. Moreover, the majority of the research has only addressed the advanced markets (e.g., the U.S. or European markets), with few studies focusing on the phenomenon in emerging economies. Therefore, the present study may shed light on how the level of the stock market crash risk in China changes during a pandemic.

## 3. Data and methodology

### 3.1. COVID-19 variables

In this study, the proxy used to measure the severity of the COVID-19 pandemic was the logarithmic growth rate of daily confirmed cases (*rCases*). We also constructed an alternative variable using the logarithmic growth rate of daily deaths (*rDths*) to run the robustness check. All of the data were retrieved from the China Stock Market & Accounting Research (CSMAR) database.

Following Da, Engelberg, and Gao (2011), we also created a COVID-19-induced fear sentiment index (*fearSent*) based on the Baidu database. In this case, if the search volume of COVID-19-related keywords was high, then it indicated that people were in fear (or even panic) about the pandemic (Salisu and Akanni, 2020). Specifically, we defined the fear index as the log of search volume plus 1. Moreover, we set a dummy variable, *D_fear*, for the fear index. In this regard, if the search volume was greater than the median of the 2020 sample, then the value of this dummy was 1 or otherwise, zero. Figure 1 displays the trends of daily confirmed cases and fear sentiment.

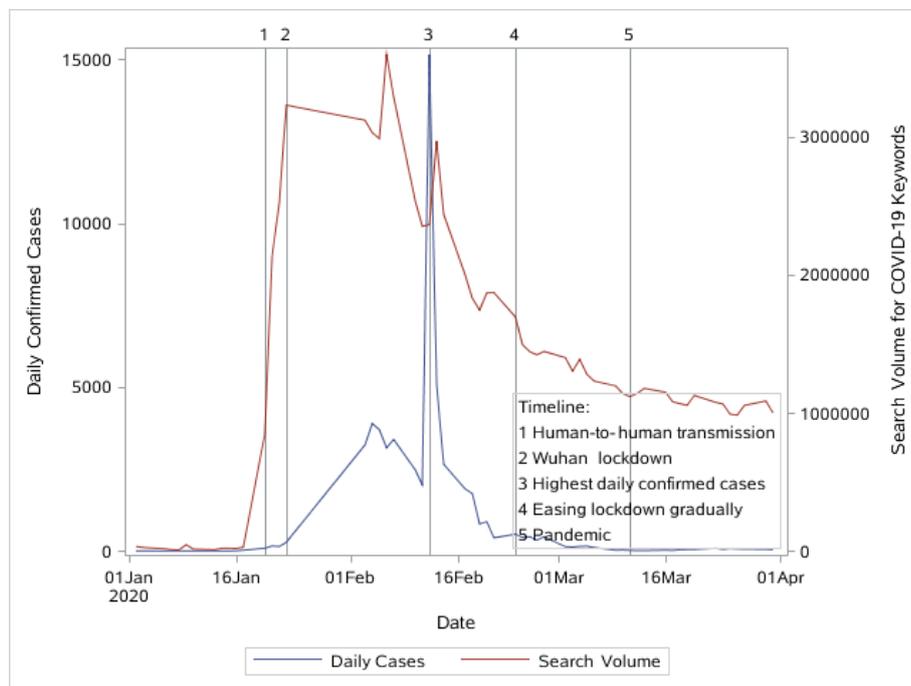

**Figure 1. The trends of daily confirmed cases and fear sentiment**

**3.2. Measuring stock market crash risk**

In this study, the market returns were collected from the value-weighted market returns of Shanghai A shares, which are frequently used in the literature (e.g., Ashraf, 2020; Al-Awadhi et al., 2020). To measure the stock market crash risk, we followed Chen, Hong, and Stein (2001), who associated such risk with the conditional skewness of the market returns. These authors calculated six-month-horizon skewness from the daily returns.

However, we used the GARCH-S (GARCH with skewness) model to estimate daily skewness. Note the following equation:

$$\begin{aligned} r_t &= \mu + \varepsilon_t; \quad \varepsilon_t \sim (0, \sigma_\varepsilon^2) \\ \varepsilon_t &= h_t^{1/2}\eta_t; \quad \eta_t \sim (0,1); \quad \varepsilon_t|I_{t-1} \sim (0, h_t) \\ h_t &= \alpha_0 + \alpha_1 \varepsilon_{t-1}^2 + \alpha_2 h_{t-1} \\ s_t &= \beta_0 + \beta_1 \eta_{t-1}^2 + \beta_2 s_{t-1} \end{aligned} \quad (1)$$

where $r_t$ is the value-weighted market returns of Shanghai A shares; $\varepsilon_t$ is the residual; $\eta_t$ is the standardized residual; $I_{t-1}$ is the information set at period $t$; $h_t$ is the conditional heteroscedasticity with a classical GARCH (1,1) structure; and $s_t$ is the conditional skewness process, which we specified as both autoregressive and dependent on lagged return shocks. In order to estimate the GARCH-S model, following Leon et al. (2005), we used a Gram–Charlier series expansion, truncated at the third moment. Note that due to the high nonlinearity of the likelihood function, we used the starting values of the parameters, estimated from the simple GARCH (1,1) model.

The market data was also obtained from the CSMAR database, with the sample period spanning from January 1, 2017 to March 31, 2020. Table 1 provides a summary of the descriptive statistics for the market returns regarding the entire sample and the subsamples, including any unconditional skewness. Note that the skewness of the entire sample was −0.71, while during the COVID-19 epidemic (January 2020–March 2020), this value was −1.47, compared to −0.30 in the 2017–2019 sample period.

| Subsample | Obs. | Mean | Min. | Max. | Std. Dev. | Skewness |
| --- | --- | --- | --- | --- | --- | --- |
| Entire sample | 789 | 0.00009 | −0.075 | 0.055 | 0.011 | −0.712 |
| Jan. 2017–Dec. 2019 | 731 | 0.0002 | −0.053 | 0.055 | 0.010 | −0.302 |
| Jan. 2020–Mar. 2020 | 58 | −0.0016 | −0.075 | 0.031 | 0.017 | −1.469 |

**Notes:** We divided our sample into two subsamples: pre- and post-COVID-19 pandemic.

**Table 1. The descriptive statistics for the market returns of the sample and subsamples**

From the statistical evidence, we did not reject the mean difference between the two subsamples in Table 1 (t-stat = 1.25, $\rho$ = 0.211), i.e., pre-COVID-19 pandemic (0.0002) and post-COVID-19 pandemic (−0.0016). This implies that there was no difference in the market returns when the COVID-19 pandemic emerged. However, with regard to the skewness index (representing market crash risk), we observed a significant difference in the mean between the two periods. More precisely, the level of the market crash in 2020 was significantly higher than that in the previous period (t-stat = 2.50, $\rho$ = 0.01). Thus, $H_1$ was not rejected, implying higher extreme volatility in the Chinese equity market during the COVID-19 pandemic.

Table 2 presents the estimation results of the GARCH-S model. As expected, there was the presence of significant conditional skewness. Specifically, the coefficient of lagged skewness was positive and significant (0.148 with t-statistic 58.079), indicating that skewness is persistent. In addition, the coefficient of the shock to skewness was positive and significant (0.036 with t-statistic 15.373), which is similar to the variance case. Overall, the majority of the coefficients were significant, implying the appropriateness of using the GARCH-S model to estimate the skewness of the market returns.

| Parameter | Value | Parameter | Value |
|---|---|---|---|
| $\mu$ | 0.00005*** | $\beta_0$ | 0.00001 |
|  | (10.47) |  | (0.98) |
| $\alpha_0$ | 0.00001*** | $\beta_1$ | 0.036*** |
|  | (39.94) |  | (15.37) |
| $\alpha_1$ | 0.117*** | $\beta_2$ | 0.148*** |
|  | (101.64) |  | (58.07) |
| $\alpha_2$ | 0.889*** | AIC | −4.556 |
|  | (891.12) |  |  |
| Log-likelihood | 1802.220 | SIC | −4.514 |

**Notes:** ***, **, and * represent statistical significance at the 1%, 5%, and 10% levels, respectively. The t-statistics are presented in parentheses.

**Table 2. Estimation results of the GARCH-S model**

Figure 2 presents the trajectory of conditional skewness, from which we can visually observe that skewness is time-varying and clustering. In particular, a significant number of cases showed negative skewness, indicating that the crash risk at these points was high. In fact, the largest negative value of skewness (−0.76) occurred during the COVID-19 outbreak, i.e., on February 4, 2020.

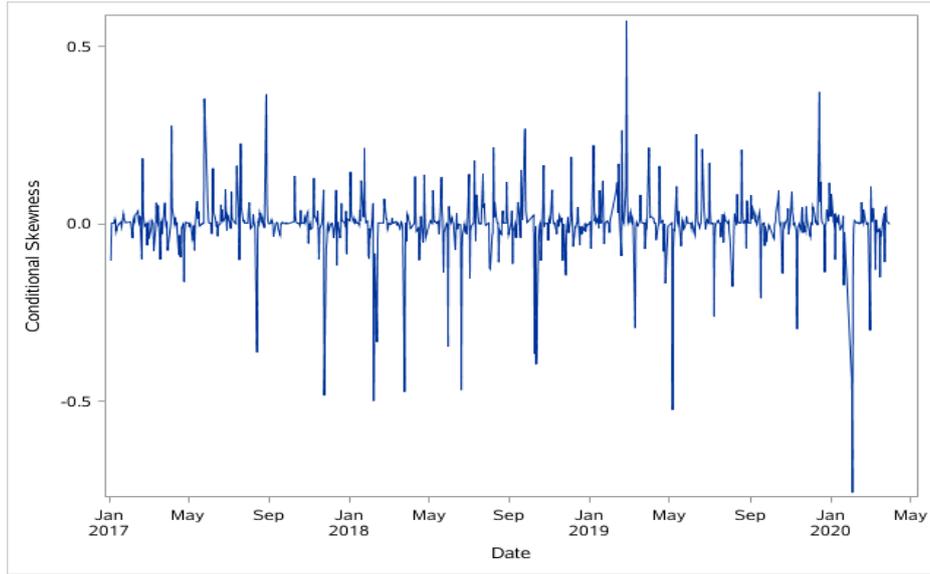

**Figure 2. Conditional skewness**

### 3.3. Model specifications

This study employed a simple time series model to examine the relationship between the COVID-19 outbreak and stock market crash risk. In this case, our dependent variable was crash risk, i.e., the conditional skewness calculated from the estimation results of the GARCH-S model. Due to the persistence of skewness, we added the lagged skewness terms in the benchmark regression model, which is specified as

$$Skew_t = c + \alpha \cdot Skew_{t-1} + \beta \cdot rCases_{t-1} + \varepsilon_t \tag{2}$$

where $Skew_t$ is the conditional skewness derived from the GARCH-S model, and $rCases$ is the logarithmic growth rate of daily confirmed cases. In addition, $c$ is a constant term, $\alpha$ and $\beta$ are the coefficients of the one-period lagged term and the logarithmic growth rate of infected cases, respectively, and $\varepsilon$ is the error term in the estimation. We denoted Equation (2) with lagged skewness, as in Model 1.

We also considered whether the COVID-19-induced fear sentiment index (*fearSent*) affects crash risk. Thus, we estimated the following model (Model 2):

$$Skew_t = c + \alpha \cdot Skew_{t-1} + \lambda \cdot fearSent_t + \varepsilon_t \tag{3}$$

$$Skew_t = c + \alpha \cdot Skew_{t-1} + \beta \cdot rCases_{t-1} + \lambda \cdot fearSent_t + \varepsilon_t \qquad (4)$$

where the terms in Equations (3) and (4) have analogous meanings to those presented above. In Equation (3), we substituted *rCase* with *fearSent* to examine how investor sentiment could predict stock market crash risk. Moreover, we used the dummy variable, *d_fear*, as the fear sentiment proxy variable to re-estimate Model 2. To further investigate the interaction effect between daily confirmed cases and fear sentiment, we added the interaction term in Model 2. In this regard, we set $\theta$ as the coefficient of the interaction term, while the other components were the same as those in Equation (4). Then, we obtained the model specification (Model 3) as follows:

$$Skew_t = c + \alpha \cdot Skew_{t-1} + \beta \cdot rCase_{t-1} + \lambda \cdot fearSent_t + \theta \cdot rCase_{t-1} \cdot fearSent_t + \varepsilon_t \qquad (5)$$

Next, we employed the Granger causality test to detect any causal relationship between stock market crash risk and fear sentiment. The model for the Granger causality test is specified as follows:

$$Skew_t = c_1 + \sum_{i=1}^{p} \alpha_i Skew_{t-i} + \sum_{j} \beta_j fearSent_{t-j} + \varepsilon_{1t} \qquad (6)$$

$$faerSent_t = c_2 + \sum_{i=1}^{p} \lambda_i Skew_{t-i} + \sum_{j=1}^{p} \delta_j fearSent_{t-j} + \varepsilon_{2t} \qquad (7)$$

where *p* is the largest lag order, which is determined through the vector autoregression model and the Bayesian information criterion. The null hypothesis for Granger causality is summarized as "*fearSent* does not cause the Granger causality to *Skew*" (*fearSent→Skew*).

For our robustness test, we selected the growth rate of daily death cases, as the substitute for the growth rate of confirmed cases, in order to predict equity market crash risk. We marked the corresponding model as Model 4, which is composed of Equations (8), (9), and (10):

$$Skew_t = c + \alpha \cdot Skew_{t-1} + \beta \cdot r(Deaths)_t + \varepsilon_t \tag{8}$$

$$Skew_t = c + \alpha \cdot Skew_{t-1} + \beta \cdot r(Deaths)_t + \lambda \cdot fearSent_{t-1} + \varepsilon_t \tag{9}$$

$$Skew_t = c + \alpha \cdot Skew_{t-1} + \beta \cdot r(Deaths)_t + \lambda \cdot fearSent_{t-1} \\ + \theta \cdot r(Deaths) \cdot fearSent_{t-1} + \varepsilon_t \tag{10}$$

Finally, because of the integration of the financial markets, we further conducted estimations to predict equity market crash risk with the number of COVID-19 infected cases and deaths. Our justification for this was that the Chinese investors not only reacted to local information but also to global news, which might have influenced their behaviors toward market crash risk. For this goal, we modified Equations (2), (4), and (5) into Equations (11), (12), and (13), respectively, to form Model 5:

$$Skew_t = c + \alpha \cdot Skew_{t-1} + \beta \cdot r(GlobalCases)_t + \varepsilon_t \tag{11}$$

$$Skew_t = c + \alpha \cdot Skew_{t-1} + \beta \cdot r(GlobalCases)_t + \lambda \cdot fearSent_{t-1} + \varepsilon_t \tag{12}$$

$$Skew_t = c + \alpha \cdot Skew_{t-1} + \beta \cdot r(GlobalCases)_t + \lambda \cdot fearSent_{t-1} \\ + \theta \cdot r(GlobalCases) \cdot fearSent_{t-1} + \varepsilon_t \tag{13}$$

## 4. Results

### 4.1. COVID-19 and stock market crash risk

The estimation results of Model 1 are reported in Table 3. In column (1), note that the coefficient of $rCases_t$ is negative and significant. This is consistent with our expectation that the COVID-19 outbreak has a negative impact on stock market crash risk. This result also reflects the reality. With the rapid spread of the pandemic, the values of listed companies were generally affected, and the stock market entered a clear economic downturn, accompanied by large declines or even crashes. Surprisingly, there was no predictive power for the other lagged terms, including $rCases_{(t-2)}$ and $rCases_{(t-3)}$ for changes in market crash risk. Hence, this finding emphasizes the role of information, particularly the number of cases, from the previous trading day on market shocks. In terms of explanatory power, the R-squared in these markets was approximately 7%, indicating that the lagged variables of the logarithmic growth rate of daily confirmed cases can be explained by the changes in market skewness, thus representing stock market crash risk.

| **Variables** | **(1)** | **(2)** | **(3)** |
|---|---|---|---|
| Intercept | −0.001 | −0.001 | −0.001 |
|  | (−0.45) | (−0.45) | (−0.45) |
| $Skew_{(t-1)}$ | 0.190*** | 0.188*** | 0.188*** |
|  | (5.53) | (5.37) | (5.35) |
| $rCases_{(t-1)}$ | −0.083*** | −0.082*** | −0.082*** |
|  | (−5.26) | (−5.17) | (−5.12) |
| $rCases_{(t-2)}$ |  | −0.005 | −0.005 |
|  |  | (−0.34) | (−0.34) |
| $rCases_{(t-3)}$ |  |  | −0.001 |
|  |  |  | (−0.05) |
| N | 787 | 786 | 785 |
| $R^2$ | 0.073 | 0.073 | 0.073 |

**Notes:** This table summarizes the estimated results for Model 1. ***, **, and * represent statistical significance at the 1%, 5%, and 10% levels, respectively. The t-statistics are presented in parentheses.

**Table 3. The effects of COVID-19 on stock market crash risk**

Our findings also confirm the findings in the literature that stock markets are more likely to be sensitive to information regarding increase in the number of confirmed cases (Albulescu, 2020; Ashraf, 2020). Thus, apart from the U.S. market, new infection cases reported at the Chinese level amplified the stock market crash risk in the country.

**4.2. Does fear sentiment matter?**

Let us now consider the role of COVID-19-induced fear sentiment. The motivation behind this is that panic about the pandemic may remain at a high level, even though the number of confirmed cases is not very large. For example, as early as January 20, 2020, academician Zhong Nanshan publicly confirmed human-to-human transmission of COVID-19 on television. Then, on January 23, the central government of China announced the lockdown of Wuhan. Although the number of confirmed cases publicly disclosed at that time was still at a relatively low level, people immediately went into a panic. In this case, the fear sentiment may have influenced the stock market before the impact from confirmed cases.

According to the results of Model 2, presented in Table 4, we used *fearSent* in the first two columns. In addition, we used a dummy variable, *D_fearSent*, to re-estimate Model 2, the results of which are shown in the last two columns. Note that all the coefficients regarding fear sentiment were negative and significant, indicating that COVID-19-induced fear sentiment can cause significant stock market crashes.

| Variables | Eq. (3) | Eq. (4) | Eq. (3-1) | Eq. (4-1) |
|---|---|---|---|---|
| Intercept | 0.045** | 0.038** | −0.0002 | 0.0001 |
|  | (2.48) | (2.13) | (−0.05) | (0.01) |
| $Skew_{(t-1)}$ | 0.188 | 0.181*** | 0.188*** | 0.181*** |
|  | (5.60) | (5.25) | (5.62) | (5.25) |
| $rCases_{(t-1)}$ |  | −0.080*** |  | −0.081*** |
|  |  | (−5.09) |  | (−5.15) |
| Fear Sentiment | −0.005*** | −0.004** | −0.042*** | −0.038** |
|  | (−2.61) | (−2.24) | (−2.61) | (−2.38) |
| N | 788 | 787 | 788 | 787 |
| $R^2$ | 0.051 | 0.079 | 0.051 | 0.080 |

**Notes:** The proxies for *Fear Sentiment* are *fearSent* in the first two columns and *D_fearSent* in the last two columns. ***, **, and * represent statistical significance at the 1%, 5%, and 10% levels, respectively. The t-statistics are presented in parentheses.

**Table 4. The effects of COVID-19-induced fear sentiment on stock market crash risk**

While Duan et al. (2020) conducted a textual analysis of 6.3 million messages on social media to conclude that the Chinese stock market most likely overreacted with growth sentiment, our findings are consistent with the aforementioned study by Da, Engelberg, and Gao (2011), who used the Baidu search engine. Interestingly, Burggraf et al. (2020) applied the same method to indicate that the Bitcoin market significantly changes when investor sentiment fluctuates. However, one of the novel points of the present study is determining where the fear sentiment stands across the pandemic. In this regard, we found that the Chinese stock market crash worsens when fear sentiment is incorporated. Our results are robust when controlling for other variables such as lagged term of skewness (the previous term for market crash risk) and the number of infected cases.

In sum, this study rejects $H_2$, which indicates that there is a relationship between investor sentiment and the stock market crash risk in China during the COVID-19 outbreak. Although the literature confirms this linkage under normal market conditions, our study sheds new light on this relationship at the onset of the pandemic.

**4.3. The interaction effect between COVID-19 and fear sentiment**

The aforementioned results show that both daily confirmed cases and fear sentiment can increase the risk of stock market crashes. This subsection further explores the inner links between these impacts and the underlying mechanisms how COVID-19 indicators could interact with fear attitudes.

Table 5 presents the results regarding the interaction effect between daily COVID-19 cases and fear sentiment. The coefficients of the interaction terms were significant and negative, indicating that fear sentiment further amplifies the negative impact of confirmed cases on stock market crash risk. In other words, fear exacerbates the negative impact of COVID-19. This highlights the importance of investors maintaining optimism during a pandemic instead of panicking about the crisis.

| Variables | Eq. (5) | Eq. (5-1) |
|---|---|---|
| Intercept | 0.036** | −0.0002 |
|  | (2.05) | (−0.07) |
| $Skew_{(t-1)}$ | 0.176*** | 0.174*** |
|  | (5.13) | (5.10) |
| $rCases_{(t-1)}$ | 0.381** | −0.0002 |
|  | (2.27) | (−0.01) |
| fearSent | −0.004** | −0.035** |
|  | (−2.17) | (−2.25) |
| $rCase_{(t-1)} \times fearSent$ | −0.032*** |  |
|  | (−2.76) |  |
| $rCase_{(t-1)} \times D\_fearSent$ |  | −0.124*** |
|  |  | (−3.77) |
| $R^2$ | 0.088 | 0.096 |

**Notes:** This table summarizes the estimated results for Model 3, including Equations (5) and (5-1). Equation (5-1) holds the dummy variable, *D_fearSent*. ***, **, and * represent statistical significance at the 1%, 5%, and 10% levels, respectively. The t-statistics are presented in parentheses. The total number of observations during this research period was 787.

**Table 5. The interaction effect between COVID-19 and fear sentiment**

It is important to consider the interaction term in our regression, for two main reasons. First, investor fear exhibits a dynamic pattern with the fatality ratio. This means that when the number of infected cases increases, investor sentiment might be affected by fear. Second, fear could mitigate risky behaviors at the onset of a pandemic, which might lead to a decrease in infected cases. Thus, examining the interaction variable constructed from the aforementioned components could offer some insight, especially on how this factor increases (or decreases) equity market crash risk.

Overall, three main conclusions can be drawn from the regression in Table 5. First, the interaction variable increases the likelihood of market crash risk at the 1% significance level. This can be explained by the fact that both factors amplify the negative impact on equity market shocks. Second, our results remained robust after substituting the fear emotion with a continuous or binary variable. This not only emphasizes a dynamic pattern but also confirms the existing role of investors' emotions in systematic risk. Third, after comparing the results in Tables 3 and 4, the explanatory level, captured by $R^2$, is substantially improved. This implies that the interaction variable may positively contribute to the explanatory feature of the changes in equity market crash risk.

However, one notable point is that as our findings mainly stemmed from the correlations between the variables, we were cautious about confirming a causal relationship before obtaining any statistical evidence. Therefore, we performed a Granger causality test to determine whether the fear emotion could increase market crash risk. Of course, the opposite direction was also examined.

## 4.4 The Granger causality test

Table 6 presents the results of the Granger causality test. In order to examine the hypothesis of each causality, we conducted an F-test. Note from Table 6 that *fearSent* is the Granger cause of *Skew*, which implies that fear sentiment causes stock market crash risk. Conversely, stock market crash risk does not Granger cause the fear sentiment.

| Direction of causality | F-test | P-value |
|---|---|---|
| fearSent → Skew | 7.1559*** | 0.0076 |
| Skew → fearSent | 1.0780 | 0.2995 |

**Notes**: ***, **, and * represent statistical significance at the 1%, 5%, and 10% levels, respectively. The null hypothesis for Granger causality is summarized as "*fearSent* does not cause the Granger causality to *Skew*" (*fearSent*→*Skew*), and the remaining hypothesis is that "*Skew* does not cause the Granger causality to *fearSent*" (*Skew* → *fearSent*).

**Table 6. The results of the Granger causality test**

Interestingly, we only observed a unidirectional Granger causality between fear sentiment and market crash risk. More precisely, fear sentiment was the factor that caused the changes in market crash risk, whereas there was no evidence in the opposite direction. Thus, we conclude that investors' attitudes toward uncertainties in terms of fear, macroeconomics, and microeconomics will stimulate stock market crash risk. Our findings also confirm the literature on fear and stock market dynamics (e.g., Bitcoin market (Chen, Liu, and Zhao, 2020), financial markets (Sharif et al., 2020), and energy markets (Salisu et al., 2020)). By examining the causal relationship, policymakers should focus on how to alleviate investor panic and maintain market stability.

**4.5 Robustness checks**

As stated earlier, an alternative proxy for measuring the severity of the COVID-19 pandemic was the growth rate of daily deaths (*rDeaths*). In this regard, we employed Model 4 to conduct our empirical analysis, the results of which are shown in Table 7. Overall, the results were consistent with our previous findings. We also substituted the number of global cases for the number of cases in China for another robustness check, as the people in China not only focused on the progress of COVID-19 at the country level but also at the global level. Table 8 presents the empirical results of Model 5.

Furthermore, we utilized global death cases to replace global confirmed cases in Model 5. As for Model 6, it comprises Equations (14), (15), and (16):

$$Skew_t = c + \alpha \cdot Skew_{t-1} + \beta \cdot r(GlobalDeaths)_t + \varepsilon_t \qquad (14)$$

$$Skew_t = c + \alpha \cdot Skew_{t-1} + \beta \cdot r(GlobalDeaths)_t + \lambda \cdot fearSent_{t-1} + \varepsilon_t \qquad (15)$$

$$Skew_t = c + \alpha \cdot Skew_{t-1} + \beta \cdot r(GlobalDeaths)_t + \lambda \cdot fearSent_{t-1} \\ + \theta \cdot r(GlobalDeaths) \cdot fearSent_{t-1} + \varepsilon_t \qquad (16)$$

Table 9 presents the empirical results of Model 6. Overall, the results shown in Tables 8 and 9 illustrate that our conclusions remained robust. Therefore, we may draw policy implications from the findings.

| Variables | Eq. (8) | Eq. (9) | Eq. (9-1) | Eq. (10) | Eq. (10-1) |
|---|---|---|---|---|---|
| Intercept | −0.001 | 0.04** | −0.0005 | 0.0357** | −0.0001 |
|  | (−0.55) | (2.29) | (−0.19) | (1.97) | (−0.05) |
| $Skew_{(t-1)}$ | 0.1873*** | 0.177*** | 0.181*** | 0.172*** | 0.172*** |
|  | (5.43) |  | (5.23) | (5.00) | (5.01) |
| $rDeaths_{(t-1)}$ | −0.115*** | −0.112*** | −0.108*** | 0.520** | 0.021 |
|  | (−5.06) | (−4.97) | (−4.70) | (2.00) | (0.53) |
| Fear Sentiment |  | −0.005** | −0.029* | −0.004** | −0.023 |
|  |  | (−2.42) | (−1.83) | (−2.08) | (−1.44) |
| $rDeaths_{(t-1)}$ × Fear Sentiment |  |  |  | −0.044** | −0.193*** |
|  |  |  |  | (−2.44) | (−3.98) |
| $R^2$ | 0.071 | 0.078 | 0.075 | 0.085 | 0.093 |

**Notes:** Table 7 summarizes the estimated results for Model 4, including Equations (8), (9), and (10). Different from Equations (9) and (10), Equations (9-1) and (10-1) hold the dummy variable, *D_fearSent*. ***, **, and * represent statistical significance at the 1%, 5%, and 10% levels, respectively. The t-statistics are presented in parentheses. The total number of observations during this research period was 787.

**Table 7. The robustness results from the number of daily deaths**

| Variables | Eq. (11) | Eq. (12) | Eq. (12-1) | Eq. (13) | Eq. (13-1) |
|---|---|---|---|---|---|
| Intercept | −0.001197 | 0.033680* | 0.000024 | 0.020837 | −0.000165 |
|  | (−0.40) | (1.83) | (0.01) | (1.12) | (−0.05) |
| $Skew_{(t-1)}$ | 0.194272*** | 0.186969*** | 0.186446*** | 0.172685*** | 0.172305*** |
|  | (5.60) | (5.37) | (5.36) | (4.98) | (4.97) |
| $rGlobalCases_{(t-1)}$ | −0.039522*** | −0.035903*** | −0.036809*** | 0.162300*** | −0.007415 |
|  | (−3.93) | (−3.51) | (−3.64) | (3.20) | (−0.59) |
| Fear Sentiment |  | −0.004065* | −0.034670** | −0.002498 | −0.023728 |
|  |  | (−1.92) | (−2.14) | (−1.17) | (−1.46) |
| $rGlobalCases_{(t-1)}$ × Fear Sentiment |  |  |  | −0.016869*** | −0.082836*** |
|  |  |  |  | (−3.99) | (−3.94) |
| $R^2$ | 0.059543 | 0.063934 | 0.065008 | 0.082579 | 0.083217 |

**Notes:** Table 8 summarizes the estimated results for Model 5, including Equations (11), (12), and (13). Different from Equations (12) and (13), Equations (12-1) and (13-1) hold the dummy variable, *D_fearSent*. ***, **, and * represent statistical significance at the 1%, 5%, and 10% levels, respectively. The t-statistics are presented in parentheses. The total number of observations during this research period was 787.

**Table 8. The robustness results from the number of daily global cases**

| Variables | Eq. (14) | Eq. (15) | Eq. (15-1) | Eq. (16) | Eq. (16-1) |
|---|---|---|---|---|---|
| Intercept | −0.001254 | 0.035187* | 0.000004 | 0.035596* | 0.000248 |
|  | (−0.42) | (1.91) | (0.00) | (1.93) | (0.08) |
| $Skew_{(t-1)}$ | 0.186324*** | 0.179297*** | 0.178773*** | 0.177722*** | 0.180314*** |
|  | (5.34) | (5.13) | (5.11) | (5.07) | (5.16) |
| $rGlobalDeaths_{(t-1)}$ | −0.049846*** | −0.045567*** | −0.046631*** | 0.137102 | −0.100354** |
|  | (−3.85) | (−3.48) | (−3.59) | (0.55) | (−1.97) |
| Fear Sentiment |  | −0.004246** | −0.035512** | −0.004297** | −0.036279** |
|  |  | (−2.01) | (−2.19) | (−2.03) | (−2.24) |
| $rGlobalDeaths_{(t-1)}$ × Fear Sentiment |  |  |  | −0.012501 | 0.057492 |
|  |  |  |  | (−0.74) | (1.09) |
| $R^2$ | 0.058829 | 0.063659 | 0.064581 | 0.064306 | 0.066005 |

**Notes:** Table 9 summarizes the estimated results for Model 6, including Equations (14), (15), and (16). Different from Equations (15) and (16), Equations (15-1) and (16-1) hold the dummy variable, *D_fearSent*. ***, **, and * represent statistical significance at the 1%, 5%, and 10% levels, respectively. The t-statistics are presented in parentheses. The total number of observations during this research period was 787.

**Table 9. The robustness results from the number of daily global deaths**

## 5. Conclusion

This study examined the relationship between the COVID-19 pandemic and the stock market crash risk in China. Based on the findings, COVID-19 increases stock market crash risk. This not only indicates that the pandemic will bring about a decline in stock market returns but that it will also aggravate the negative symmetry of stock market returns and will increase the possibility of extreme downturns in stock prices. We also found that even when the number of confirmed cases is not significantly large, people's fears about the virus will increase stock market crash risk. Finally, we found that fear sentiment not only directly increases crash risk but may also boost the negative impact of COVID-19 on stock market crash risk.

Overall, this study is a reminder that preventing panic during a pandemic is helpful for reducing stock market crash risk. Thus, we draw two main policy implications. First, closer observation by lawmakers of the financial markets with regard to the dynamics of fear and the number of cases is necessary. In this regard, regulators should determine how to immediately and effectively support the market when fear is overwhelming. By doing so, market crash risk can be managed in extreme cases. Second, investors are not only likely to be sensitive to local information (i.e., Chinese domestic infected cases or deaths) but also to global news. Therefore, clear and timely communication regarding the COVID-19 pandemic could bring about effective prediction in the market. More importantly, both investors and regulators should be more cautious about stock market crash risk when the number of cases (or deaths) significantly increases. Then, hedging or safe haven strategies could be implemented, as suggested by previous research (Conlon et al., 2020; Conlon and McGee, 2020).